# An Effective Dereverberation Algorithm by Fusing MVDR and MCLP

*Fengqi Tan, Changchun Bao*\*

Speech and Audio Signal Processing Laboratory, Faculty of Information Technology,
Beijing University of Technology, Beijing, 100124, China
fengqi0712@emails.bjut.edu.cn, baochch@bjut.edu.cn

**Abstract**

In the scenario with reverberation, the experience of human-machine interaction will become worse. In order to solve this problem, many methods for the dereverberation have emerged. At present, how to update the parameters of the Kalman filter in the existing dereverberation methods based on multichannel linear prediction (MCLP) is a challenging task, especially, accurate power spectral density (PSD) estimation of target speech. In this paper, minimum variance distortionless response (MVDR) beamformer and MCLP are effectively fused in the dereverberation, where the PSD of target speech used for Kalman filter is modified in the MCLP. In order to construct a MVDR beamformer, the PSD of late reverberation and the PSD of the noise are estimated simultaneously by the blocking-based PSD estimator. Thus, the PSD of target speech used for Kalman filter can be obtained by subtracting the PSD of late reverberation and the PSD of the noise from the PSD of observation signal. Compared to the reference methods, the proposed method shows an outstanding performance.

## 1. Introduction

Reverberation and noise often cause performance degradation in many speech processing applications. In particular, the presence of reverberation seriously affects the accuracy of sound source localization and results in the distortion of speech signal [1]. With fast development of speech communication, the interaction between people and smart devices is becoming more and more frequent and important [2], [3]. How to suppress reverberation in noisy environment for improving the performance of various speech devices is still a problem that has attracted a lot of attention.

Currently, for most of dereverberation methods, the power spectrum of target speech signal is very important to the performance of the algorithm. For example, the well-known weighted prediction error (WPE) [4] aims at iteratively updating the linear prediction filter coefficients and the power spectral density (PSD) of target signal to obtain optimal late reverberation prediction. However, the traditional WPE method needs to process long-term speech signal for obtaining more accurate PSD estimation of target signal, so it cannot process speech in real time. Therefore, the online-WPE based methods [5-7] that utilize PSD estimation of target signal were proposed. Minimum variance distortionless response (MVDR) beamforming [8] requires the estimation of covariance matrix of the noise and the steering vector of target signal relative to the reference microphone to obtain the optimal spatial filter coefficients, so the estimation of covariance matrix of the noise becomes the key to improve algorithm performance. The covariance matrix of noise is usually obtained by the microphone signals without voice activity [9, 10]. However, in high reverberation scenario, it is very difficult to obtain the covariance matrix of late reverberation, so how to calculate filter coefficients of MVDR beamformer is challenge. In order to solve the problem that the noise and reverberation occur simultaneously, some methods combining beamformer and multi-channel linear prediction have also been proposed [11-14]. Specially, the method integrating sidelobe cancellation and linear prediction (ISCLP) [15] can finish the task of denoising, dereverberation and cancelling interference at the same time. This method used Kalman filter to simultaneously estimate the coefficients of sidelobe cancellation filter and linear prediction filter. In recent years, some estimation methods of the power spectral of late reverberation and noise were proposed successively [16-18], such as PSD estimation of late reverberation based on neural network [16] and the PSD estimation of late reverberation and noise by maximum likelihood (ML) [18]. These PSD estimation methods have been proved to be effective by the experiments given in [19] and [20].

In this paper, we propose an effective dereverberation algorithm by reasonably fusing MVDR beamformer and linear predictor based on Kalman filtering. In this algorithm, the covariance matrixes of late reverberation and noise are estimated by the same method given in [17]. The blocking-based PSD estimator is employed to construct the MVDR beamformer. The late reverberation is modeled by the multichannel linear prediction model, where the linear prediction coefficients are obtained by the Kalman filter derived by the state equation and the observation equation.

The rest of the paper is arranged as follows. In Section 2, the signal model used in this paper is defined. In Section 3, the blocking-based PSD estimator are introduced briefly. In Section 4, the proposed dereverberation algorithm combined the blocking-based PSD estimator, MVDR and Kalman filter are described. In Section 5, the experimental simulation is carried out and the experimental results are analyzed. The conclusions are given in Section 6.

## 2. Signal Model

We assume that there are $M$ microphones collecting speech signals in a room with the reverberation and noise. The collected signal of the $m^{th}$ microphone is expressed as $y_m(k,l)$ in terms of short-time Fourier transform (STFT), where $k$ and $l$ denote the frequency bin and frame index, respectively. In frequency-domain, the stacked vector of the signals collected by microphones is denoted as

$$\mathbf{y}(k,l)=[y_1(k,l),\cdots,y_M(k,l)]^T \qquad (1)$$

where T indicates transpose of the matrix. In the cases of reverberation and noise, the stacked vector of the corrupted signals collected by microphones can be written as

$$\mathbf{y}(k,l) = \mathbf{x}_e(k,l) + \mathbf{x}_l(k,l) + \mathbf{v}(k,l) \quad (2)$$

where the vector $\mathbf{x}_e(k,l)$ denotes the STFT of direct speech signal with early reverberation, it is given by

$$\mathbf{x}_e(k,l) = q(k,l)\mathbf{d}(k) \quad (3)$$

where $q(k,l)$ is target signal and $\mathbf{d}(k)=[d_1(k),\cdots,d_M(k)]^T$ is the relative transfer function (RTF) vector of target signal between the reference microphone and all microphones. The vector $\mathbf{x}_l(k,l)$ denotes the STFT of late reverberation signals, and vector $\mathbf{v}(k,l)$ denotes the STFT of noise signal. In this paper, the noise is regarded as spatially uncorrelated. In the following contents, since signal processing is independently operated in each frequency bin, the index of frequency bin is omitted for simplification.

## 3. Blocking-based PSD Estimator

Blocking-based PSD estimator was proposed in [17], which aims to block target speech signal and estimate covariance matrixes of late reverberation and noise. The blocking matrix $\mathbf{B}$ is constructed such that

$$\mathbf{B}^H\mathbf{d}=0 \quad (4)$$

where $\mathbf{d}$ is the aforementioned RTF vector of target signal. The signals $\mathbf{n}(l)$ containing noise and late reverberation at the $l^{th}$ frame can be obtained multiplying the collected signals $\mathbf{y}(l)$ by the blocking matrix, i.e

$$\mathbf{n}(l)=\mathbf{B}^H\mathbf{y}(l) \quad (5)$$

The covariance matrix of $\mathbf{n}(l)$ at the $l^{th}$ frame is obtained as follows:

$$\mathbf{\Phi}_n(l) = E\{\mathbf{n}(l)\mathbf{n}^H(l)\}=\phi_r(l)\underbrace{\mathbf{B}^H\mathbf{\Gamma}\mathbf{B}}_{\tilde{\mathbf{\Gamma}}} + \phi_v(l)\underbrace{\mathbf{B}^H\mathbf{\Psi}\mathbf{B}}_{\tilde{\mathbf{\Psi}}} \quad (6)$$

where the symbol E indicates expectation operator, $\phi_r(l)$ is the PSD of late reverberation at the $l^{th}$ frame, $\mathbf{\Gamma}$ is the spatial coherence matrix of diffuse sound field, which can be obtained by the microphone array geometry. $\phi_v(l)$ is the PSD of noise, $\mathbf{\Psi}$ is the spatial coherence matrix of the noise. When the noise is spatially uncorrelated, $\mathbf{\Psi}$ can be regarded as an identity matrix. By solving the following matrix equation, the PSD of late reverberation and noise can be obtained

$$\begin{bmatrix} \mathrm{tr}\{\tilde{\mathbf{\Gamma}}^H\tilde{\mathbf{\Gamma}}\} & \mathrm{tr}\{\tilde{\mathbf{\Gamma}}^H\tilde{\mathbf{\Psi}}\} \\ \mathrm{tr}\{\tilde{\mathbf{\Gamma}}^H\tilde{\mathbf{\Psi}}\} & \mathrm{tr}\{\tilde{\mathbf{\Psi}}^H\tilde{\mathbf{\Psi}}\} \end{bmatrix} \begin{bmatrix} \phi_r(l) \\ \phi_v(l) \end{bmatrix} = \begin{bmatrix} \mathrm{tr}\{\hat{\mathbf{\Phi}}_n^H(l)\tilde{\mathbf{\Gamma}}\} \\ \mathrm{tr}\{\hat{\mathbf{\Phi}}_n^H(l)\tilde{\mathbf{\Psi}}\} \end{bmatrix} \quad (7)$$

With the covariance matrices of late reverberation and noise, the signals collected by microphones can be processed by MVDR beamformer for obtaining the enhanced signal, where the PSD of target signal is obtained as

$$\phi(l) = \frac{1}{\mathbf{d}^H\mathbf{d}}\mathrm{tr}\{\mathbf{\Phi}_\mathbf{y} - \phi_r(l)\mathbf{\Gamma} - \phi_v(l)\mathbf{\Psi}\} \quad (8)$$

## 4. Proposed Algorithm

The block diagram of the proposed dereverberation method is shown in Fig. 1, in which there are three paths. The upper one performs the MVDR beamforming based on the blocking-based PSD estimator to obtain the enhanced signal $x_b(l)$. The PSD $\phi(l)$ of target signal used for Kalman filter also can be obtained by the blocking-based PSD estimator. The middle one performs the blocking-based PSD estimator. The lower one performs multichannel linear prediction based on Kalman filter to obtain predictive signal, i.e., reverberation signal $r(l)$ by using the delayed signals $\mathbf{t}(l)$ of input signal $\mathbf{y}(l)$. The finally enhanced signal $s(l)$ is obtained by subtracting $r(l)$ from $x_b(l)$. Different from the ISCLP structure [15], the MVDR beamformer is used to replace GSC in this paper, because the GSC beamforming has the defect of distorting speech signal, that is, if the steering vector of the GSC cannot be estimated accurately, its block matrix will damage speech elements inevitably.

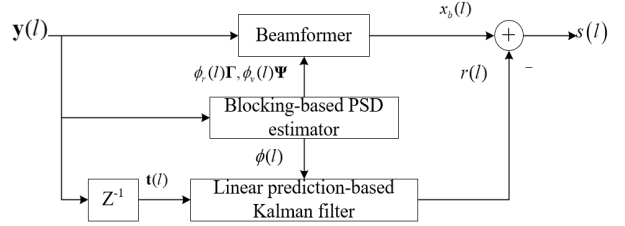

Fig. 1: *Block diagram of the proposed dereverberation method*

### 4.1. Multichannel linear prediction-based Kalman filter

In the upper path of Fig. 1, the enhanced speech signal $x_b(l)$ after beamformer can be expressed as

$$x_b(l) = \mathbf{w}_b^H(l)\mathbf{y}(l) \quad (9)$$

where $\mathbf{w}_b^H(l)$ is the coefficients of MVDR beamformer [17], i.e.

$$\mathbf{w}_b(l) = \frac{\left[\phi_r(l)\mathbf{\Gamma} + \phi_v(l)\mathbf{\Psi}\right]^{-1}\mathbf{d}}{\mathbf{d}^H\left[\phi_r(l)\mathbf{\Gamma} + \phi_v(l)\mathbf{\Psi}\right]^{-1}\mathbf{d}} \quad (10)$$

Assuming that the uncorrelated noise is largely removed by MVDR, thus, the speech signals processed by beamformer are composed of late reverberation and direct signal with early reverberation. In the lower path of Fig. 1, the input of linear predictor is composed of the signals from previous (L-D) frames. By stacking vector $\mathbf{y}(l)$, the input matrix of the $l^{th}$ frame of linear predictor is expressed as

$$\mathbf{t}(l)=\left[\mathbf{y}^T(l-D),\cdots,\mathbf{y}^T(l-L+1)\right]^T \quad (11)$$

and the predicted signal $r(l)$, i.e., the late reverberation signal, can be expressed as

$$r(l) = \mathbf{w}_b^H(l)\mathbf{x}_l(l)=\sum_{n=D}^{L}\mathbf{w}_b^H(l)\mathbf{G}_{r,n}(l)\mathbf{y}(l-n) \\ =\mathbf{w}_r^H(l)\mathbf{t}(l) \quad (12)$$

where $\mathbf{G}_{r,n}(l)$ is the multichannel linear prediction coefficients, $\mathbf{w}_r^H(l)$ is the stacked vector of $\mathbf{w}_b^H(l)\mathbf{G}_{r,n}(l)$. By selecting an appropriate $D$ according to the overlap of adjacent frames in the STFT processing, we can make direct and early reflection signal $\mathbf{x}_e(l)$ and late reverberation signal $\mathbf{x}_l(l)$ uncorrelated, i.e.

$$\langle \mathbf{x}_e(l), \mathbf{x}_l(l) \rangle = \sum_l \mathbf{x}_e(l)\mathbf{x}_l(l) = 0 \quad (13)$$

Considering the architecture of entire algorithm, the output, i.e., the finally enhanced signal, is obtained by

$$s(l) = x_b(l) - \mathbf{w}_r^H(l)\mathbf{t}(l) \quad (14)$$

where $s(l)$ can be regarded as the estimation of $t(l)$.

Kalman filter has been used very well for estimating linear prediction coefficients [21]. Based on Kalman filtering

principle, in this paper, the observation equation of Kalman filter is defined as

$$x_b^*(l) = \mathbf{t}^H(l)\mathbf{w}_r(l) + x_e^*(l) \qquad (15)$$

where the symbol "*" denotes the conjugation operation, $\mathbf{w}_r(l)$ is called state vector with zero mean, and its related covariance matrix is

$$\mathbf{\Phi_w}(l) = E\left[\mathbf{w}_r(l)\mathbf{w}_r^H(l)\right] \qquad (16)$$

$x_e^*(l)$ in (15) is the estimation of target signal, also called measurement error. The state equation is defined using the first-order Markov process as follows

$$\mathbf{w}_r(l) = \mathbf{A}^H(l)\mathbf{w}_r(l\text{-}1) + \mathbf{v_w}(l) \qquad (17)$$

where $\mathbf{A}(l)$ is state transition matrix that represents the prediction of state vector from previous frames to current frame, and process noise $\mathbf{v_w}(l)$ is modeled by zero-mean Gaussian model with the following covariance matrix

$$\mathbf{\Phi_v}(l) = E\left[\mathbf{v_w}(l)\mathbf{v_w}^H(l)\right] \qquad (18)$$

According to the state equation and observation equation, the update equations and gain $\mathbf{k}(l)$ of Kalman filter [15] can be obtained as follows:

$$\hat{\mathbf{w}}_r(l) = \mathbf{A}(l)\hat{\mathbf{w}}_r^+(l\text{-}1) \qquad (19)$$

$$\mathbf{\Phi}_e(l) = \mathbf{A}^H(l)\mathbf{\Phi}_e^+(l\text{-}1)\mathbf{A}(l) + \mathbf{\Phi_v}(l) \qquad (20)$$

$$s^*(l) = x_b^*(l) - \mathbf{t}^H(l)\hat{\mathbf{w}}_r(l) \qquad (21)$$

$$\phi_s(l) = \mathbf{t}^H(l)\mathbf{\Phi}_e(l)\mathbf{t}(l) + \phi_{x_e}(l) \qquad (22)$$

$$\mathbf{k}(l) = \mathbf{\Phi}_e(l)\mathbf{t}(l)\phi_s^{-1}(l) \qquad (23)$$

$$\hat{\mathbf{w}}_r^+(l) = \hat{\mathbf{w}}_r(l) + \mathbf{k}(l)s^*(l) \qquad (24)$$

$$\mathbf{\Phi}_e^+(l) = \mathbf{\Phi}_e(l) - \mathbf{k}(l)\mathbf{t}^H(l)\mathbf{\Phi}_e(l) \qquad (25)$$

where $\mathbf{\Phi}_e(l)$ is the covariance matrix of state error, $\phi_{x_e}(l)$ is the PSD of target signal.

### 4.2. PSD estimation of target speech

As described in Section 3, the covariance matrix of late reverberation and noise can be obtained as $\phi_r(l)\mathbf{\Gamma} + \phi_v(l)\mathbf{\Psi}$, the PSD of target speech that is necessary for Kalman filter can be obtained by equation (8). In this paper, the PSD of target speech is estimated by

$$\phi_{x_e}(l) = \alpha\phi(l) + (1-\alpha)\left|\mathbf{w}_b^H(l\text{-}1)\mathbf{y}(l)\right|^2 \qquad (26)$$

where $0 < \alpha < 1$ is the weighting factor. In equation (6), noise and reverberation are modeled by coherence matrix, the error of the coherence matrix is inevitable, so the power spectral density estimator is not completely reliable. The goal of the weighting factor is to adjust the PSD of target signal to ensure that PSD of target signal is in a reliable range.

In this way, the blocking-based PSD estimator can work for MVDR beamformer and Kalman filter at the same time, can calculate the covariance matrix of late reverberation and noise from the received signals of the microphones, and can also calculate the PSD estimation of the target signal. The steering vector in blocking-based PSD estimator can be obtained by the direction of arrival (DOA) of speech source and the microphone array geometry.

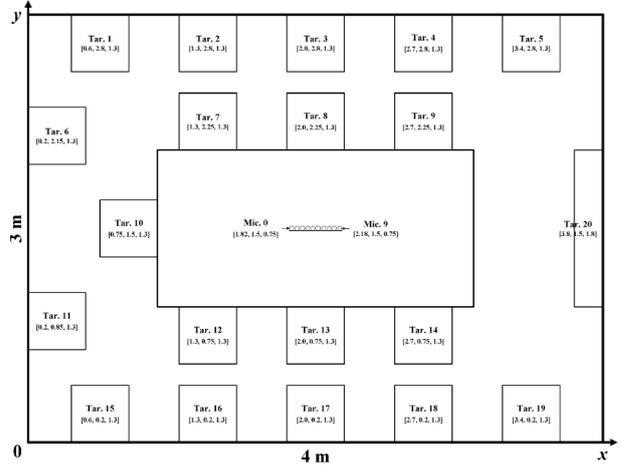

Fig. 2: *The room acoustic environment of microphone array speech simulator, in which there are 20 source locations and 10 microphones are placed on the middle of a table.*

## 5. Experiments

In this section, the proposed method will be compared with the MVDR beamforming with blocking-based PSD estimator [17] and ISCLP [15].

### 5.1. Experimental setup

The speech signals required for the experiment are generated by the microphone array speech simulator of the indoor acoustic environment proposed in [22], and the speech dataset comes from the TIMIT [23]. As shown in the Fig. 2, a linear microphone array is placed in the center of the room, where 10 microphones with 4cm inter-microphone distance collect speech signals from all angles in the room. We select 8 microphones for our experiments. One source position, namely Tar.1 are chosen for the test, and 100 utterances are selected. For building reverberate speech with noise, babble noise is used at 5dB and 10dB input SNR levels. The reverberation time T60 is set to 400ms, 500ms, 600ms, 700ms and 800ms, respectively. We assume that the generated speech signals contain background noise of the microphones, and still applied to the current signal model. The speech signals are sampled at 16 kHz. The length of Hanning window with 50% overlap is 512 samples. A 512-point STFT is used for the windowed speech signals. The delay $D$ that preserves the early reflections is set to 2, and the order $L$ of linear prediction filter is set to 10. The steering vector for beamforming and blocking matrix is determined by the direction angle of source [15].

In order to evaluate the proposed algorithm, we choose the perceptual evaluation of speech quality (PESQ) [24], the short-time objective intelligibility (STOI) [25] and the signal-to-interference ratio (SIR) [26] as the measures.

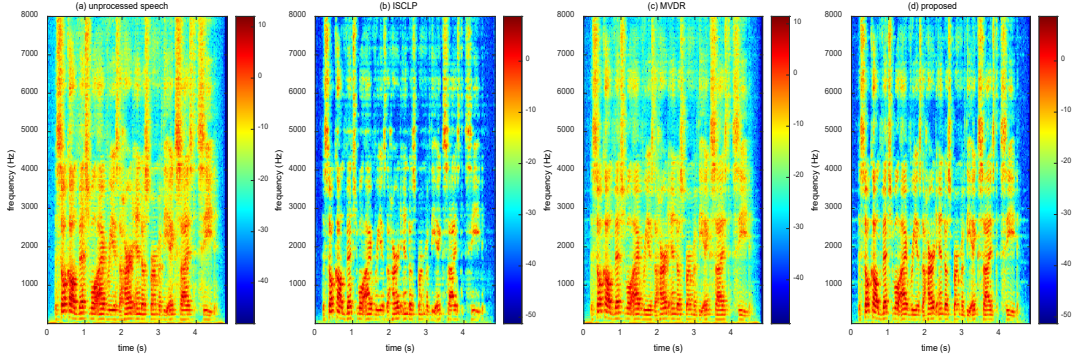

Fig. 3: *Spectrogram comparison at the conditions of $T_{60}=800ms$.*

Table 1: *Test results for 5dB babble noise*

| $T_{60}$ [ms] | 400 | 500 | 600 | 700 | 800 |
|---|---|---|---|---|---|
| PESQ | | | | | |
| unprocessed | 2.16 | 2.15 | 2.10 | 2.08 | 2.06 |
| ISCLP | **2.38** | 2.30 | 2.22 | 2.20 | 2.18 |
| MVDR | 2.31 | 2.33 | 2.22 | 2.21 | 2.20 |
| proposed | 2.34 | **2.36** | **2.28** | **2.31** | **2.33** |
| STOI | | | | | |
| unprocessed | 0.73 | 0.72 | 0.70 | 0.69 | 0.67 |
| ISCLP | 0.75 | 0.74 | 0.73 | 0.73 | 0.72 |
| MVDR | 0.79 | **0.78** | **0.76** | 0.75 | 0.75 |
| proposed | **0.80** | **0.78** | **0.76** | **0.76** | **0.77** |
| SIR | | | | | |
| unprocessed | -4.90 | -5.90 | -6.59 | -7.11 | -7.51 |
| ISCLP | -0.84 | -1.15 | **-1.12** | **-1.36** | **-1.85** |
| MVDR | -2.18 | -3.09 | -4.73 | -5.12 | -5.46 |
| proposed | **-0.11** | **-0.77** | -1.59 | -1.97 | -2.13 |

Table 2: *Test results for 10dB babble noise*

| $T_{60}$ [ms] | 400 | 500 | 600 | 700 | 800 |
|---|---|---|---|---|---|
| PESQ | | | | | |
| unprocessed | 2.38 | 2.31 | 2.24 | 2.19 | 2.15 |
| ISCLP | 2.56 | 2.47 | 2.40 | 2.32 | 2.30 |
| MVDR | 2.51 | 2.34 | 2.39 | 2.25 | 2.21 |
| proposed | **2.56** | 2.43 | **2.53** | **2.51** | **2.50** |
| STOI | | | | | |
| unprocessed | 0.77 | 0.75 | 0.73 | 0.71 | 0.69 |
| ISCLP | 0.78 | 0.77 | 0.76 | 0.75 | 0.75 |
| MVDR | 0.80 | 0.79 | 0.77 | 0.76 | 0.76 |
| proposed | **0.81** | **0.79** | **0.78** | **0.77** | **0.77** |
| SIR | | | | | |
| unprocessed | -4.91 | -5.91 | -6.60 | -7.12 | -7.54 |
| ISCLP | -0.81 | **-1.09** | **-1.02** | **-0.96** | **-0.62** |
| MVDR | -2.19 | -3.52 | -3.91 | -4.25 | -4.45 |
| proposed | **-0.06** | -1.43 | -2.11 | -1.78 | -1.85 |

### 5.2. Experimental results

Fig. 3 shows an example of spectrogram comparison, in which Fig. 3(a) ~ Fig. 3(d) correspond to the unprocessed speech signal of reference microphone, the processed speech signals by ISCLP, MVDR and the proposed algorithm. It can be seen that, compared with the unprocessed speech signal, ISCLP and proposed method effectively reduce the reverberation, but MVDR does not work well for dereverberation. Among them, the ISCLP suppressed a lot of reverberation, but the distortion of the target signal is also serious, i.e., a lot of frequency information is lost at high frequencies, which makes the speech sound less bright. The proposed algorithm also reserves the components of high frequency meanwhile reducing reverberation effectively. Compared with the ISCLP, the proposed algorithm shows much clearer spectral structure. In the listening perception, the proposed algorithm produced more acceptable perception.

For evaluation, two kinds of input SNR levels and five kinds of reverberation time are considered. Tables 1 and Table 2 show the results of performance evaluation of the four algorithms at 5dB and 10dB SNR level, respectively. It can be seen that the PESQ and STOI of the proposed algorithm are higher than two reference algorithms in most cases, and the STOI is significantly improved especially. For the SIR, ISCLP shows the best effect in most cases. Within a certain distortion range, the proposed method can remove more interference than the traditional MVDR beamforming. In the future work, the proposed method needs to be improved in terms of reducing speech distortion.

### 6. Conclusions

In this paper, an effective algorithm for the dereverberation was proposed by combining MVDR beamforming based on blocking-based PSD estimator and multichannel linear prediction with the Kalman filter. The covariance matrix of late reverberation and noise for constructing MVDR beamforming was obtained by the blocking-based PSD estimator, and the PSD of target signal was also be obtained by the PSD estimator, which is an important parameter for Kalman filter. The evaluation results proved that the proposed algorithm reduced reverberation remarkably and improved speech quality. The application of PSD estimator in both beamforming and linear prediction improved the computational efficiency and achieved well performance.

### 7. Acknowledgements

This work was supported by the National Natural Science Foundation of China (Grant No.61831019).